\documentclass[a4paper]{aa} % for a referee version
\usepackage{graphicx}
\usepackage{natbib}
\usepackage{times}
\def\subsun{\mbox{$_{\normalsize\odot}$}}
\def\lesssim{\mathrel{\hbox{\rlap{\hbox{\lower4pt\hbox{$\sim$}}}\hbox{$<$}}}}
\def\lya{\mbox{Ly$\alpha$}}

\def\ecs{\mbox{~erg~cm$^{-2}$~s$^{-1}$}}

\def\gtrsim{\mathrel{\hbox{\rlap{\hbox{\lower4pt\hbox{$\sim$}}}\hbox{$>$}}}}

\begin{document}
\title{Extended Lyman-$\alpha$ emission around bright quasars
\thanks{Based on observations obtained at the German-Spanish
  Astronomical Center, Calar Alto, operated by the Max-Planck-Institut
  f\"ur Astronomie Heidelberg jointly with the Spanish National
  Commission for Astronomy. } }

\author{L. Christensen \inst{1,2} \and K. Jahnke \inst{2,3} \and L. Wisotzki
  \inst{2} \and S.~F. S\'anchez \inst{4}}
     
   \institute{European Southern Observatory, Casilla 19001, Santiago 19, Chile, \email{lichrist@eso.org}
    \and  Astrophysikalisches Institut Potsdam, An der Sternwarte 16,
     14482 Potsdam, Germany
    \and  Max Planck Institut f\"ur Astronomie, K\"onigstuhl 17, 69117
Heidelberg, Germany
    \and Centro Astronomico Hispano Aleman de Calar Alto, Calle Jesus Durb\'an
    Rem\'on 2,2 E-04004 Almeria, Spain}

%    \mail{L. Christensen} 
%  \titlerunning{Extended \lya\ emission around QSOs}
   \date{Received ; accepted}

%________________________________________________________________
   
   \abstract 
% context heading (optional) 
{Quasars trace the most massive structures at high redshifts and their
presence may influence the evolution of the massive host galaxies.}
% aims heading(mandatory)
{We study the extended \lya\ emission line regions (EELRs) around
seven bright, mostly radio-quiet quasars (QSOs) at $2.7<z<4.5$, and
compare luminosities with EELRs around radio-loud QSOs reported in the
literature.}
% methods heading (mandatory) 
{ Using integral field spectroscopy, we analyse the morphology and
  kinematics of the quiescent \lya\ EELRs around the QSOs. }
% results heading (mandatory) 
{ We find evidence for the presence of EELRs around four radio-quiet
  and one radio-loud QSO. All EELRs appear asymmetric and the
  optically brightest QSOs also have the brightest \lya\ nebulae.  For
  the two brightest nebulae we find velocities between
  $\sim$600~km~s$^{-1}$ at the QSO position to $\sim$200~km~s$^{-1}$
  at a distance of $3-4$\arcsec\ from the QSO and surface flux
  densities up to 2--3$\times10^{-16}$ \ecs arcsec$^{-2}$. The five
  EELRs have total \lya\ luminosities which correspond to $\sim$0.5\%
  of the luminosities from the QSOs broad \lya\ emission lines. This
  fraction is an order of magnitude smaller than found for EELRs
  around radio-loud, steep spectrum QSOs reported in the
  literature. While the nebulae luminosities are correlated with the
  QSO \lya\ luminosities, we find that nebulae luminosities are not
  correlated with the central QSO ionising fluxes.}
% conclusions heading (optional)
{ The presence of gas in the EELRs can be interpreted based on two
  competing scenarios: either from quasar feedback mechanisms, or from
  infalling matter.  Apart from these two effects, the \lya\ flux
  around radio-loud objects can be enhanced due to interactions with
  the radio jets. The relatively fainter nebulae around radio-quiet
  QSOs compared to lobe-dominated radio-loud QSOs can be ascribed to
  this effect, or to significant differences in the environments
  between the two classes.}


   \maketitle   

%________________________________________________________________
\section{Introduction}

Models of galaxy evolution require feedback mechanisms, e.g. from
supernovae to regulate the star formation in order to reproduce galaxy
luminosity functions \citep[e.g.][]{white91}.  Recent simulations
\citep{dimatteo05,springel05} and semi-analytical models
\citep{croton05} have indicated that also a central active galactic
nucleus can have strong effects on the surroundings. Such strong
feedback mechanisms can be tested observationally by studying the
environment of bright high redshift quasars, specifically by
observations of the kinematics of extended gas.

In this paper we focus on observations of extended \lya\ emission line
regions (EELRs) surrounding $z>2$ QSOs. \lya\ EELRs are frequently
found for radio-loud quasars (RLQ) \citep{heckman91}. The
\lya\ emission nebulae are mostly asymmetric around the lobe-dominated
RLQs \citep{heckman91b,lehnert98}, and the nebulae are aligned with
the spatial orientation of the radio jets. Furthermore, these studies
showed that the strongest line emission is mostly associated with the
location of the stronger radio lobe.  This alignment effect indicates
some interaction of the radio emission and the EELRs through
shocks. Radio galaxies, believed to contain obscured QSOs with the
radio jet oriented closer to the plane on the sky than regular QSOs
\citep{barthel89}, are also found to have EELRs aligned with the radio
emission \citep{mccarthy95}. Besides these EELRs, which are possibly
powered by jet interactions, radio galaxies at $z>2$ are found to have
more extended and quiescent \lya\ halos with velocities around 500
km~s$^{-1}$ \citep{villar-martin03}.

EELRs have also been reported around radio-quiet quasars
(RQQs\footnote{In this paper the term QSO is used as a broad term that
  combines all types of quasars in one category. When the radio flux
  is specifically used to distinguish between radio-loud and quiet
  QSOs, the term RLQ or RQQ will be used.}), typically based on
long-slit spectra of a few individual objects
\citep{steidel91,bremer92,fried98,moller00,bunker03,weidinger04}, or
based on narrow-band images \citep{bergeron99,fynbo00b}.  The EELRs
around RQQs need to be explained by effects other than interactions
with jets, and possible explanations include the ionising radiation
from massive stars in the host galaxy or from the AGN itself. Although
the majority of quasars are radio-quiet, the phenomenon of extended
emission around RQQs at high redshifts has not been studied
systematically. From a larger sample of 12 QSOs, \citet{hu87} only
found extended emission from one object, which suggests that the
phenomenon is not common.  Other investigations have reported
companion \lya\ emitting objects at the QSO redshifts that may be
influenced by the QSO ionising radiation \citep{hu96,petitjean96}.

Apart from the feedback effect which will cause an outflow of the
surrounding material, infalling matter also can be responsible for the
observed \lya\ photons.  In a scenario where matter is falling into a
potential well, \lya\ photons are created via cooling processes, and
ionisation by a QSO can greatly enhance the brightness of the EELRs.
In a model of infalling material, \citet{haiman01} predict that
\lya\ emission should be detectable in a region of
$\sim$3\arcsec\ around QSOs, equivalent to $\sim$25 kpc at
$z\approx3$, and with a typical surface flux density of
$10^{-16}-10^{-18}$ \ecs
arcsec$^{-2}$. \citet{weidinger04,weidinger05} found observational
support for this interpretation by suggesting that the emission from
the QSO is directed in a cone. However, studies of spatially extended
\lya\ emission line regions have failed to answer the question whether
the emitting gas is infalling or outflowing.

This paper presents a study of seven QSOs at $2.7<z<4.5$ to look
  for narrow ($\lesssim$1000 km~s$^{-1}$) \lya\ emission lines at the
QSO redshifts. Because the spatial location of \lya\ emission at the
QSO redshift is not known in advance and can be highly asymmetric,
integral field spectroscopy (IFS) is useful to locate the brightest
EELRs.  Previous IFS studies of EELRs around RLQs have focused on
optical emission lines from low redshift objects
\citep[e.g.][]{durret94,crawford00,sanchez04b,christensen06}, or the
analysis of a single radio galaxy \citep{villar05}.  IFS is well
suited for this purpose because of the possibility to create
narrow-band images at any suitable wavelength with an adjustable band
width, and to simultaneously investigate velocity profiles.

This paper is organised as follows. Sect.~\ref{sect:qlya_dataset}
describes the data reduction, and Sect.~\ref{sect:analysis} the
analysis of the IFS data, with more detailed information on each
object and the velocity structure of the EELRs. We explore scaling
relations between the EELRs and the QSO luminosities in
Sect.~\ref{sect:qlya_scaling}, and discuss these in
Sect.~\ref{sect:qlya_disc}. The conclusions are presented in
Sect.~\ref{sect:conc}. Throughout the paper, we assume a flat
cosmology with $\Omega_{\Lambda}=0.7$ and $H_0=70$ km~s$^{-1}$
Mpc$^{-1}$.

%========================================================================
%========================================================================
\section{Data set}
\label{sect:qlya_dataset}
The observations were performed with the Potsdam Multi Aperture
Spectrophotometer \citep[PMAS; ][]{pmas00,roth05}, mounted on the 3.5m
telescope at Calar Alto during several observing runs in the period
2002--2004. The study was carried out in connection with a different
programme to study emission from Damped \lya\ absorbers
\citep[e.g.][]{christensen04}.  The sample consists of seven objects
where the QSO \lya\ line lies within the spectral window.  The
redshifts of the damped systems were different to the QSO redshifts
and therefore \lya\ emission lines detected at the QSO redshifts are
clearly related to the QSO environments. No attempts were made to
construct a complete or unbiased sample with respect to QSO
properties, and the sample contained one flat spectrum, core-dominated
RLQ and six RQQs at $2.7<z<4.5$ as listed in
Table~\ref{tab:qlya_list_obj}.  Furthermore, only observations of the
4 lowest redshift objects covered the wavelength of \ion{C}{iv}
$\lambda$1549 at the QSO redshift.  Because of varying conditions, the
total integration times for each object were adjusted to give the same
signal-to-noise levels in the data cubes. A log of the observations is
given in Table~\ref{tab:log}.

\begin{table*}
\begin{footnotesize}
\centering
  \begin{tabular}{llllllll}
   \hline \hline
   \noalign{\smallskip}
 (1)& (2) &(3) & (4) & (5) & (6) & (7) &(8)\\
  Coordinate name & Alt. name & $z_{\mathrm{em}}$ (ref) & $z_{\mathrm{em}}$ (cube) &  mag. (cube)  & seeing & exposure time (s) & class\\
   \hline
   \noalign{\smallskip}
\object{Q0953+4749} & \object{PC 0953+4749}& 4.457&   --  & $R$=20.0
   &1.0 & 16200 & RQQ  \\
   \noalign{\smallskip}
\object{Q1347+112}&                        & 2.679& 2.679 & $V$=18.7 &   0.6
   & 12600 & RQQ\\
   \noalign{\smallskip}
\object{Q1425+606}& \object{SBS 1425+606}  & 3.17 & 3.203 & $V$=16.5 & 1.0 &
   10800 & RQQ \\
   \noalign{\smallskip}
\object{Q1451+1223}& \object{B1451+123}    & 3.246& 3.261 & $V$=18.6 &  0.8
   & 12600  & RQQ\\
   \noalign{\smallskip}
\object{Q1759+7539}& \object{GB2 1759+7539}& 3.050 & 3.049  & $V$=17.0
   &1.2 & 12600 & RLQ \\
   \noalign{\smallskip}
\object{Q1802+5616}&\object{PSS J1802+5616}& 4.158&-- & $R$=20.7  &
   1.0 & 27000 & RQQ \\
   \noalign{\smallskip}
\object{Q2233+131}&         & 3.298&--  & $V$=18.3  & 0.7 & 18000 &  RQQ \\
   \noalign{\smallskip}
   \hline
   \noalign{\smallskip}
  \end{tabular}
  \caption[]{List of observed QSOs. Redshifts for the QSOs reported in
   the literature (Column 3) are compared to the redshift determined
   from the IFS data cube (Column 4).  The '--' signs correspond to
   data where we could not reliably determine the QSO redshift from
   the emission lines in the IFS data cubes. Redshifts are derived
   from the vacuum corrected wavelengths. Column 5 gives the QSO
   magnitudes in the Vega system, and column 6 the seeing in
   arcseconds measured in the data cubes.  Column 7 lists the total
   integration time and the last column is the class of object defined
   from radio observations.  }
  \label{tab:qlya_list_obj}
\end{footnotesize}
\end{table*}

\begin{table*}
\begin{footnotesize}
\centering
  \begin{tabular}{lllllll}
   \hline \hline
   \noalign{\smallskip}
QSO & date & exposure time & grating & $\lambda$ coverage & seeing & conditions\\
    &      & (s)   & & ({\AA})\\
  \noalign{\smallskip}
   \hline
   \noalign{\smallskip}
Q0953+4749     & 2004-04-16 & 4$\times$1800  & V300 & 3630--6980 & 0.9 & stable\\
               & 2004-04-21 & 5$\times$1800  & V300 & 3630--6980 & 1.0 & non phot.\\
Q1347+112      & 2004-04-20 & 7$\times$1800  & V300 & 3630--6980 & 0.6 & non phot.\\
Q1425+606      & 2004-04-19 & 6$\times$1800  & V300 & 3630--6750 & 1.0 & stable\\
Q1451+1223     & 2004-04-17 & 7$\times$1800  & V300 & 3630--6980 & 0.8 & non phot.\\
Q1759+7539     & 2004-04-21 & 7$\times$1800  & V300 & 3630--6980 & 1.0--1.5
   & non phot. \\
Q1802+5616     & 2003-06-18 & 2$\times$1800  & V600 & 5100--6650 & 1.0 & non phot.\\ 
               & 2003-06-20 & 3$\times$1800  & V600 & & 1.0 & non phot. \\
               & 2003-06-21 & 4$\times$1800  & V600 & & 1.8 & non phot.\\
               & 2003-06-22 & 6$\times$1800  & V600 & & 0.9 & stable\\
Q2233+131      & 2003-08-24 & 6$\times$1800  & V600 & 4000--5600 &  0.6 & stable\\
                & 2003-08-25 & 4$\times$1800  & V600 & & 0.7 & non phot.\\
   \noalign{\smallskip}
   \hline
  \end{tabular}
  \caption[]{Log of the observations. The last two columns show the
   average seeing during the integrations and the photometric
   conditions, respectively.}
  \label{tab:log}
\end{footnotesize}
\end{table*}

A detailed description of the data reduction and analysis is presented
along with the Damped \lya\ absorber study \citep[][; Christensen et
  al., 2006, in prep]{christensen05}.  In summary, after the data
reduction each observation of the QSO was contained in a data cube of
spatial dimensions of 8\arcsec$\times$8\arcsec with a contiguous
sampling. The data cube had 16$\times$16 spatial elements
(``spaxels'') where each spaxel covered a square of 0\farcs5 on a
side.  The spectral range covered was from 1500 to 3000~{\AA},
and a spectral resolution of 3 and 6 {\AA} was obtained for the two
setups used.  The wavelength calibration was done in the standard
manner using exposures of emission line lamps.  Errors in the
wavelength calibration in the final data cubes were estimated from
strong sky emission lines indicating typical uncertainties of 0.3
{\AA}. Flux calibrations used standard procedures by observations of
spectrophotometric standard stars at the beginning and the end of each
night.

%________________________________________________________________
\section{Analysis}
\label{sect:analysis}
This section describes the processes to detect EELRs in the data
cubes, and notes for each individual object. A more detailed analysis
of the two brightest nebulae is also presented.

\subsection{Subtracting the QSO emission}

Since the QSO continuum and broad-line nuclear \lya\ emission was by
far the dominating contribution to the overall luminosity at
rest-frame wavelength of 1215~{\AA}, a subtraction of the nuclear
emission was essential in order to search for a possible quiescent
Ly$\alpha$ EELR around the nucleus.  To give an idea of the extension
of the PSF, the broad emission lines could be found even at a radial
distance of 3\farcs5 from the centers of the brightest QSOs.  A visual
inspection of the data cubes without the nuclear contribution
subtracted gave no clear indication of the presence of such EELRs.
Extended continuum emission from stars in the host galaxies is not
expected to be measurable at this angular resolution
\citep[c.f.][]{jahnke04}. Three different methods for subtracting the
QSO emission were investigated.

%\subsubsection{On-off band subtraction}
The simplest method to study EELRs uses conventional narrow-band
filters to observe an emission line and the continuum adjacent to the
line. Exactly the same procedure can be used for the data cubes, with
the advantage that we are free to choose a suitable wavelength range
for both the on- and off bands. Since the redshift of quiescent
\lya\ emission was previously unknown, the central wavelengths and
widths for the on-band images were determined from the emission lines
found using the two other methods described below.  The spectral range
used to produce the offband images were the same as for the on-band
images, but redwards by about 10 {\AA}. This ensured that the nuclear
line and continuum emission was removed, and produced narrow-band
images of the quiescent component of the EELRs.  The left hand panels
in Fig.~\ref{fig:qlya_qso_lya} show such residual narrow-band images
centered on the wavelengths where narrow \lya\ emission lines were
found.  The peak intensities were determined by fitting
two-dimensional Gaussian functions in the on- and off-band images.
Then the off-band image was scaled to the on-band peak and subtracted
to create pure EELR images.

%\subsubsection{Scaling the QSO emission}
\label{sect:qlya_psf}
Another method takes advantage of the fact that the spectrum of a
point source is the same, although scaled, in adjacent spaxels.  From
each reduced data cube a one-dimensional QSO spectrum was created by
co-adding spectra from several fibres within a 2\arcsec\ radial
aperture centered on the QSO. For each spaxel we determined a scale
factor between the total one-dimensional QSO spectrum and each
individual spaxel in an interval bracketing the QSO \lya\ emission
line and subtracted it from the initial spectrum.  To a first
approximation this process created a data cube that retained the EELR.
To refine the process of finding pure emission lines, this residual
emission line cube was subtracted from the original cube to create a
cube with the QSO emission only.  A new one-dimensional QSO spectrum
was created, scaled to each spaxel and subtracted, again from the
original data cube.  The process was iterated until stable solution
was found which occurred usually after 3 iterations only. After these
3 iterations, the total flux of the EELRs extracted from the residual
spectra  was found not to change with further iterations.

In the residual data cube, we analysed the spectra in the EELR and
determined the spectral \textit{FWHM} of the emission lines shown in
the left hand panels in Fig.~\ref{fig:qlya_qso_lya}.  To find these
widths, one-dimensional spectra associated with the emission lines
were created first. Then images of the EELRs were created by choosing
band widths corresponding to twice the measured \textit{FWHM} of the
emission lines and centered on the \lya\ emission wavelength.  More
adjacent spaxels were added to the one-dimensional spectra if they
showed up as bright points in the narrow-band images. This process was
iterated interactively using the Euro3D visualisation tool
\citep{sanchez04a}.  Typically for the fainter objects, apertures of
1\farcs5 in diameter was used, which correspond to about 10 spaxels.
For the brighter objects (Q1425+606 and Q1759+7539) a much larger
aperture was used, and the one-dimensional spectra present a sum of 60
spaxels.

\begin{figure*}[!htpb]
\centering
\resizebox{15cm}{!}{\includegraphics[bb= 37 100 535 826, clip]{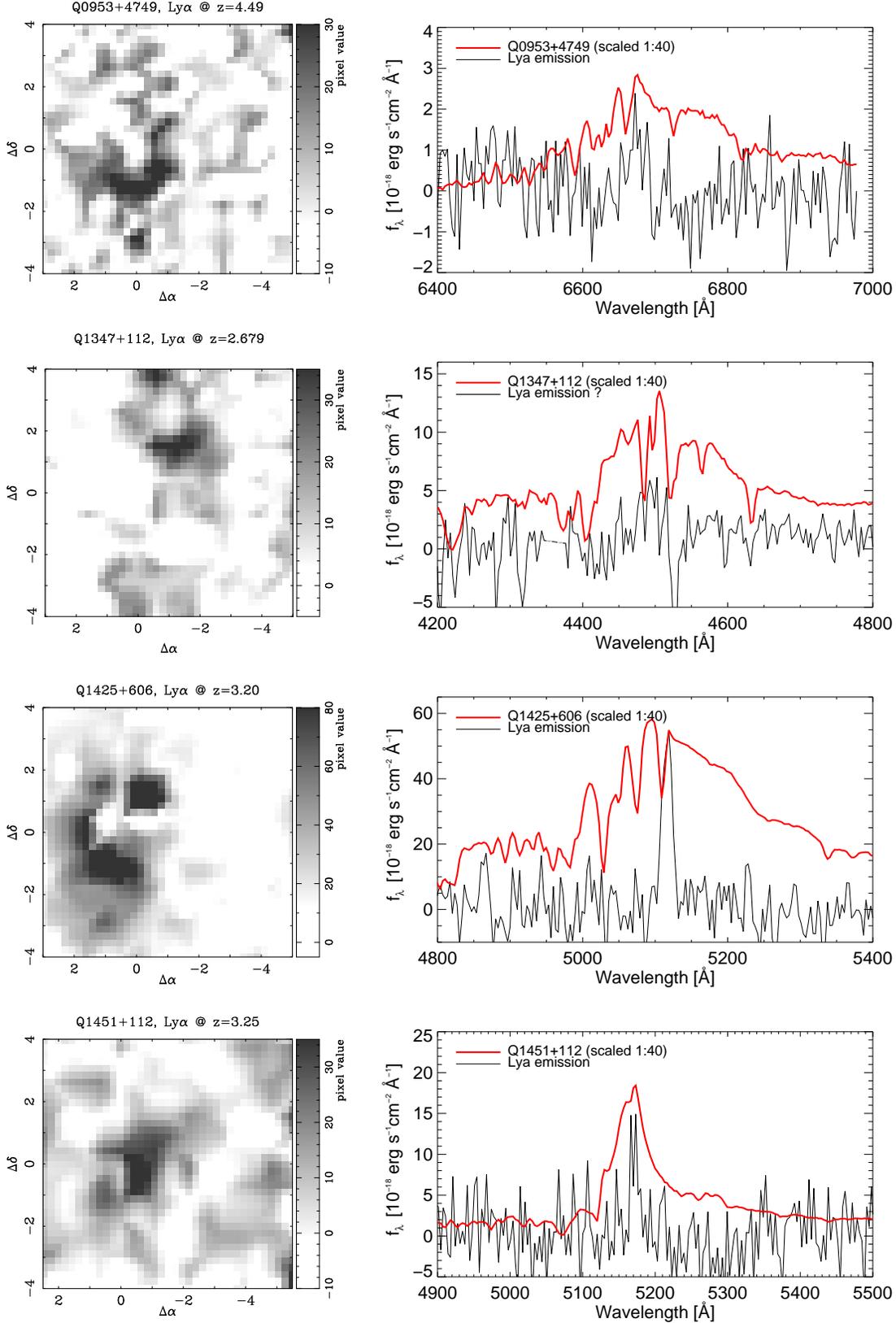}}
\caption{Extended \lya\ emission line regions at the QSO
  redshifts. \textit{The left hand panels} show
  8\arcsec$\times$8\arcsec\ narrow-band images created using a simple
  on-band minus off-band image technique. The centers of the QSOs are
  placed at the coordinates (0,0). All images are interpolated
  representations with pixel sizes of 0\farcs2 with the orientation
  north up and east left.  The bright objects (Q1425+606 and
    Q1759+7539) have strong residuals within the central
  1\arcsec. The \textit{Right hand panels} show the spectra of the
  extended narrow \lya\ emission (thin line) compared with the QSO
  spectra scaled down by a factor of 40 (thick line). The spectrum of
  Q1802+5616 was only scaled down by a factor of 3. No extended
  emission is found from Q1802+5616 and Q1347+112. Where found, the
  \lya\ emission lines from the EELRs are clearly narrower than the
  broad QSO lines.}
\label{fig:qlya_qso_lya}
\end{figure*}
\addtocounter{figure}{-1}
\begin{figure*}[!t]
\centering   
\resizebox{15cm}{!}{\includegraphics[bb= 37 270 535 826, clip]{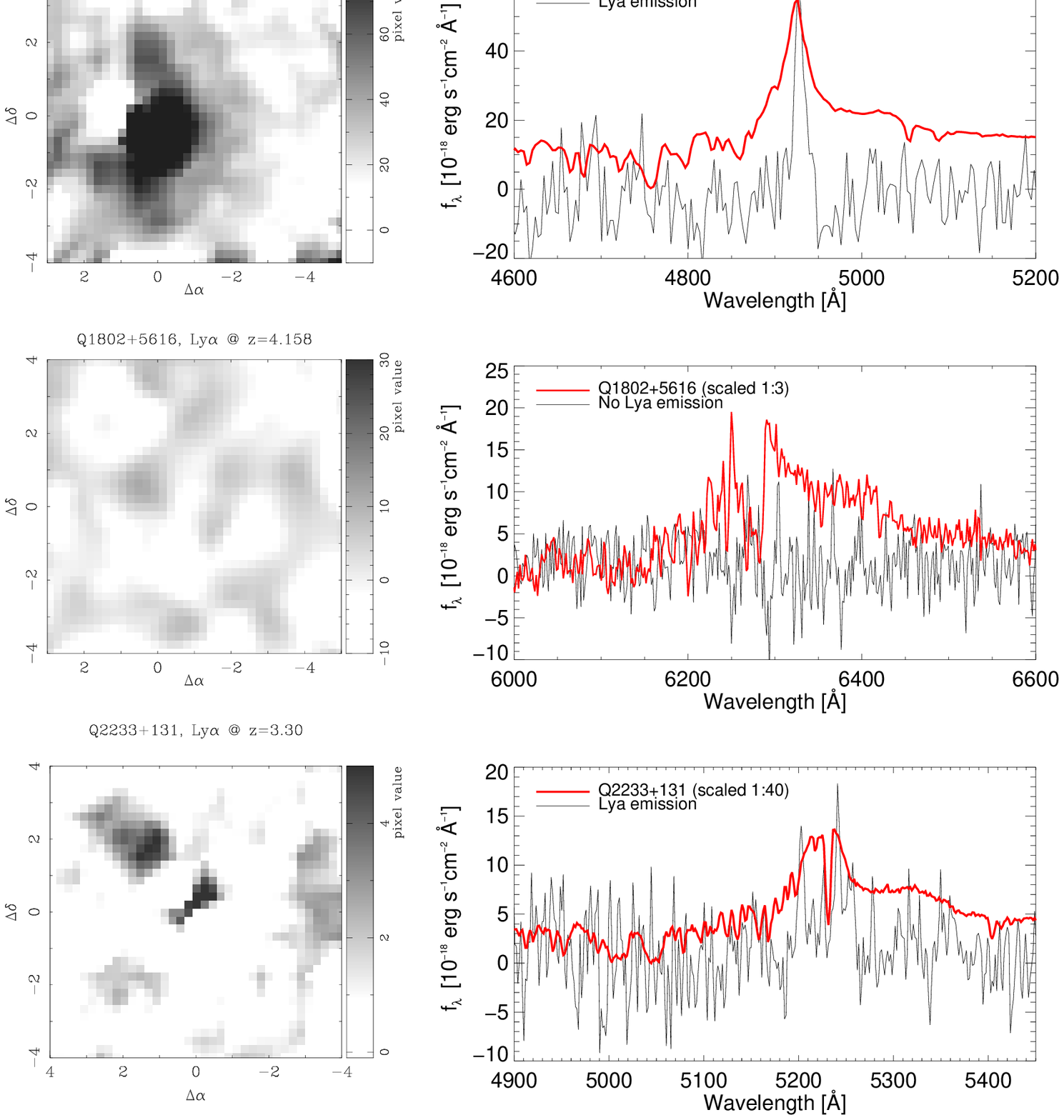}}
\caption{\textit{continued.}}
\end{figure*}

%\subsubsection{PSF modeling}
The third method used a technique of modeling the two-dimensional QSO
image PSF in each monochromatic slice
\citep[e.g.][]{wisotzki03,sanchez04b,sanchez06}.  First, a
two-dimensional model of the QSO PSF is made for each monochromatic
slice in the data cube.  By fitting two-dimensional Gaussian functions
to these images we determine the spatial location, and the
\textit{FWHM} in x- and y directions of the QSO emission. Since these
parameters vary smoothly with wavelength, we can use this information
to make a model PSF in the form of a data cube. This model data cube
is subtracted from the original data to create a residual data cube.

%\subsubsection{Comparison of methods}
Qualitatively, the three methods gave the same impression about the
presence and morphology of the EELRs, but each method had its
advantages and drawbacks. The on-off band subtraction procedure was
used to create the images in Fig.~\ref{fig:qlya_qso_lya}, but since
this procedure accounted for one band only, we could not use this
technique to analyse the spectra associated with the emission, which
is the true advantage of IFS. Therefore, this method was only used to
verify that the images derived from the more complex procedures were
reproduced.

Comparing the results from the PSF scaling and the spectral scaling
methods we found that the spectral properties, e.g.  emission line
fluxes and widths, were similar, but that significantly larger QSO
residuals were present for the first method. Hence we decided to use
the spectral scaling method for further analysis, unless noted
otherwise. The difference between the two methods will be demonstrated
in Sect.~\ref{sect:q1759}. A probable reason for the larger residuals
from the PSF subtraction was the fact that the PSF could not
adequately be described by a simple analytical function.  On the other
hand, when we analysed one single object (i.e. a point source) with a
well known spectrum similar in every single spaxel, the spectral
scaling method provided better results. However, in the case that the
EELR have strong, point source like emission directly in the line of
sight to the QSO, we probably over-subtracted the extended
emission. In such a case, none of the procedures would be able to
reconstruct the emission.

%________________________________________________________________
\subsection{Notes on individual objects}
\label{sect:qlya_notes}

In this section the results are presented in detail for each
individual object, and much of the description refers to the images
and spectra in Fig.~\ref{fig:qlya_qso_lya}.  Non-photometric
conditions during the observations or intrinsic variability of the
QSOs make it difficult to determine whether spectrophotometric results
were obtained. The absolute flux calibration uncertainties are
estimated to be $\sim$20\% on the basis on comparing the
spectrophotometry with broad band magnitudes for the quasars reported
in the literature. We cannot disentangle any absolute magnitude
differences from the intrinsic flux variability of the QSOs though.

\subsubsection{Q0953+4749}
Only the QSO \lya\ and Ly$\beta$ emission lines are included in the
spectral window and both are affected heavily by absorption lines in
the \lya\ forest.  Hence no attempt is made to estimate the QSO
redshift from the data cube. 
 
The on-band image in Fig.~\ref{fig:qlya_qso_lya} for Q0953+4749
corresponds to 6656--6682 {\AA}.  We find extended narrow
\lya\ emission at 6670~{\AA} mostly towards the south of the QSO.
This wavelength is redshifted by $\sim$1000 km~s$^{-1}$ relative to
the QSO redshift. The existence of asymmetric emission extending over
5\arcsec\ was found in a deep long-slit spectrum by \citet{bunker03}.
The signal-to-noise level in the data cube is too low to investigate
the dynamics and morphology of this emission line region or to trace
the emission over 5\arcsec.  Apparently the brighter part of the
emission in Fig.~\ref{fig:qlya_qso_lya} extends over more than
2\arcsec\ in the data cube.

\subsubsection{Q1347+112}

 The narrow-band image for Q1347+112 corresponds to 4466--4482~{\AA}.
 \lya\ emission from this QSO appears in a region of the CCD affected
 by a bad column, which after QSO subtraction causes a distinct
 structure extending over the field in a narrow-band image. This can
 be seen to the north-west of the QSO center in
 Fig.~\ref{fig:qlya_qso_lya}. The emission line coincides with a sky
 emission line from which sky-subtraction residuals can be interpreted
 as narrow emission lines. Nonetheless, faint extended emission
 appears to be present, but further analysis is not possible.

\subsubsection{Q1425+606}

 The narrow-band image for Q1425+606 corresponds to 5099--5122 {\AA}.
 To estimate the QSO redshift a Gaussian function is fit to the
 \ion{O}{i} $\lambda$1302 emission line with
 $z_{\mathrm{em}}=3.2030\pm0.0008$.  This is somewhat higher than
 $z=3.17$ reported in the literature, which we also derive from the
 \ion{C}{ii} $\lambda$1335 and \ion{C}{iv} $\lambda$1549 lines in the
 data cube. As a reference we use $z_{\mathrm{em}}=3.203$ because the
 \ion{O}{i} line is a more reliable redshift indicator
 \citep{tytler92}.

Q1425+606 has one of the brightest EELR in the sample, as shown in
Fig.~\ref{fig:qlya_qso_lya}, and it is also the brightest QSO in the
sample.  It has large QSO subtraction residuals, but these are at a
level consistent with Poissonian statistics.  Any narrow
\lya\ emission that may be present within a radius of 0\farcs5 from
the QSO center is heavily affected by these subtraction residuals,
hence we do not recover the structure of the nebula here.

The \lya\ emission is clearly asymmetric with most of the emission
coming from a region to the south-east of the QSO.  Co-adding spaxels
in the region where the narrow \lya\ emission is brightest gives a
spectrum with a \lya\ emission line with a total line flux of
$(7\pm1)\times10^{-16}$ \ecs.

Subtracting the QSO emission around the \ion{C}{iv} line reveals no
emission line in either the narrow-band image or one-dimensional
spectrum. We derive a 3$\sigma$ upper detection limit of
6$\times$10$^{-17}$ \ecs.

\subsubsection{Q1451+112}
 The narrow-band image for Q1451+112 corresponds to 5163--5173 {\AA}.
 The broad QSO \lya\ emission line is strongly affected by absorption
 lines in the \lya\ forest. An estimate of the QSO redshift is
 difficult because the red wing of the \ion{O}{i} $\lambda$1302 line
 is affected by residuals from the 5577 {\AA} sky line. However, we
 adopt $z=3.261$ estimated from this emission line. A lower redshift
 of $z=3.2468$ is inferred from the \ion{C}{iv} $\lambda$1549 line.

After QSO spectral subtraction a faint narrow emission line region is
found extending to $\sim$2\arcsec\ to the north-west and west of the
QSO as seen in Fig.~\ref{fig:qlya_qso_lya}. This narrow emission line
has $z=3.2530$ implying a velocity difference of +700 km~s$^{-1}$
relative to \ion{C}{iv} from the QSO broad line region but a blue
shift of --600 km~s$^{-1}$ relative to the \ion{O}{i} line.  Further
analysis of the structure of the emission line region is not possible
with the signal-to-noise level in the individual spaxels.

\subsubsection{Q1759+7539}
\label{sect:q1759}
 The narrow-band image for Q1759+7539 corresponds to 4919--4933 {\AA}.
 This object is the only RLQ in the sample with a core-dominated,
 flat-spectrum radio emission \citep{hook96}, and like Q1425+606 it is
 also very bright in the optical.  A Gaussian fit to the QSO
 \ion{C}{iv} emission line gives $z=3.0486\pm0.0008$, where the error
 includes the wavelength calibration error.  The \lya\ emission line
 gives a similar result; other emission lines are fainter or do not
 allow for a good determination of the redshift because absorption
 lines affect the line profiles. The on--off band image in
 Fig.~\ref{fig:qlya_qso_lya} shows the clear presence of an extended
 structure to the south-west of the QSO center.

\begin{figure*}
\centering  
\resizebox{15cm}{!}{\includegraphics[bb= 37 640 535 826,clip]{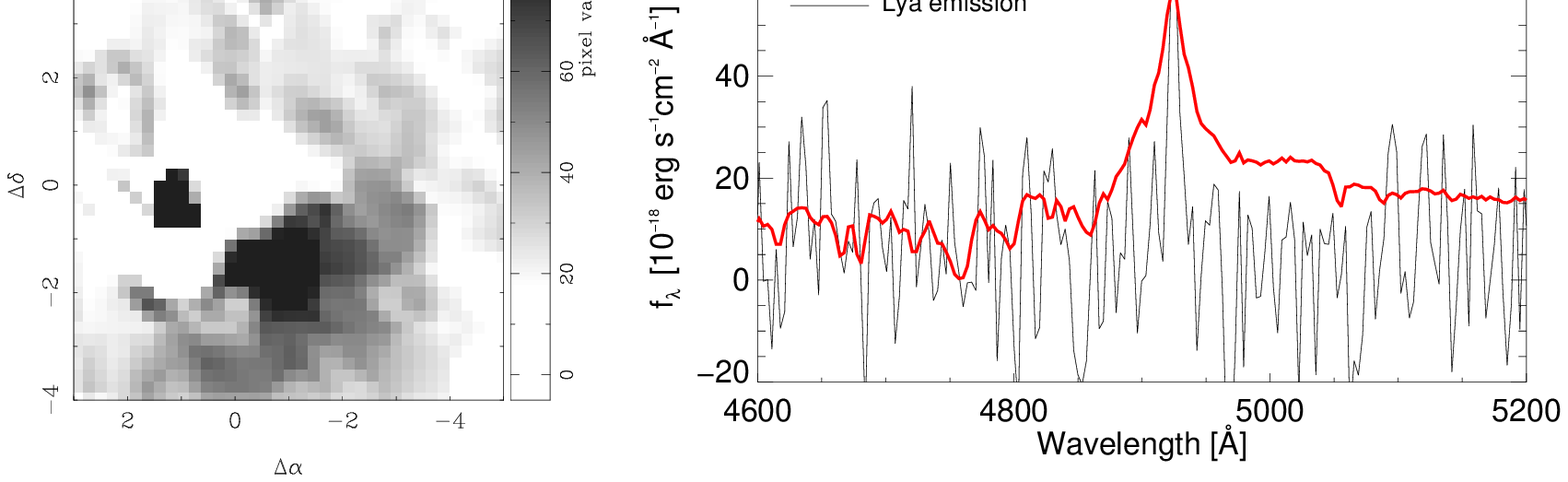}}
\caption{Similar to Fig.~\ref{fig:qlya_qso_lya}, but using an
alternative QSO subtraction method, where the PSF is modeled as a
function of wavelength. The image still shows strong residuals close
to the QSO center. Compared to the spectrum in
Fig.~\ref{fig:qlya_qso_lya}, the one-dimensional spectrum of the EELR
is more noisy.}
\label{fig:q17_spec_version2}
\end{figure*}

The two different methods for the QSO PSF subtraction were
investigated in more detail. Modeling the QSO PSF as a function of
wavelength gives a narrow-band image very similar to that in
Fig.~\ref{fig:qlya_qso_lya}. However, we find that the spectrum is
much noisier as demonstrated in Fig.~\ref{fig:q17_spec_version2}.
Most importantly, we find that the total flux in the extracted
one-dimensional spectrum remains the same within the uncertainties.

As for Q1425+606, we do not find extended \ion{C}{iv} emission from
this object either,  with an upper limit of 9$\times$10$^{-17}$
  \ecs.

\subsubsection{Q1802+5616}

 The narrow-band image for Q1802+5616 corresponds to 6270--6277 {\AA}.
 No \lya\ emission line nebula is found at the QSO redshift, but this
 is possibly because the QSO itself if fainter than the others in this
 sample. The QSOs continuum emission is $\sim$40 times fainter than
 for the two bright QSOs in our sample. Thus if any correlation
 between the QSO and EELR flux is present, the narrow \lya\ EELR
 should have a total flux of 3$\times10^{-17}$ \ecs\ which is below
 the detection limit in this data cube. This flux has been estimated
 applying a global scaling relation between the QSO emission and the
 EELR flux as detailed in Sect.~\ref{sect:qlya_scaling}. Because the
 QSO \lya\ emission line falls close to the strong sky emission line
 at 6300 {\AA} and is affected by strong residuals, this object is not
 considered further here.

\subsubsection{Q2233+131}
 The narrow-band image for Q2233+131 corresponds to 5223--5230 {\AA}.
 Only the broad \lya\ emission line lies within the spectral range in
 the data cube, and this line only is insufficient to estimate the QSO
 redshift.  The broad \lya\ line is affected by both \lya\ forest
 lines and a strong \ion{Si}{II} $\lambda$1260 absorption line
 associated with a DLA system \citep[see][]{christensen04}.  Since no
 adequate redshift could be determined from the data cube, we rely on
 the literature value for the QSO redshift.

After QSO subtraction we find an emission line object 1\arcsec\ to the
north-east of the QSO.  The scaled spectrum subtraction method reveals
a region that appears to have an emission line at 5227 {\AA} which is
at the same wavelength as the \ion{Si}{ii} absorption line in the QSO
spectrum. In contrast the PSF subtracted cube also shows an emission
region to the north-east, but the associated emission line is at 5240
{\AA}, i.e. shifted by 700 km~s$^{-1}$ relative to the QSO
redshift. In the case of the scaled spectrum subtraction, an
inaccurate wavelength calibration could introduce a shift in the
wavelength of the emission line due to the presence of the absorption
line, while this is not the case for the PSF subtraction.  Therefore
Fig.~\ref{fig:qlya_qso_lya} and Table~\ref{tab:qlya_results} present
the results from the PSF subtraction technique. To the limit of
detection this region does not appear to be very extended.

\subsection{Velocity structure}
\label{sect:vel}
We now proceed with a more detailed investigation of the velocity
structures and morphologies for the two brightest nebulae.

\begin{figure*}[!htpb]
\centering
\resizebox{\hsize}{!}{\includegraphics[bb= 40 400 555 795,clip]{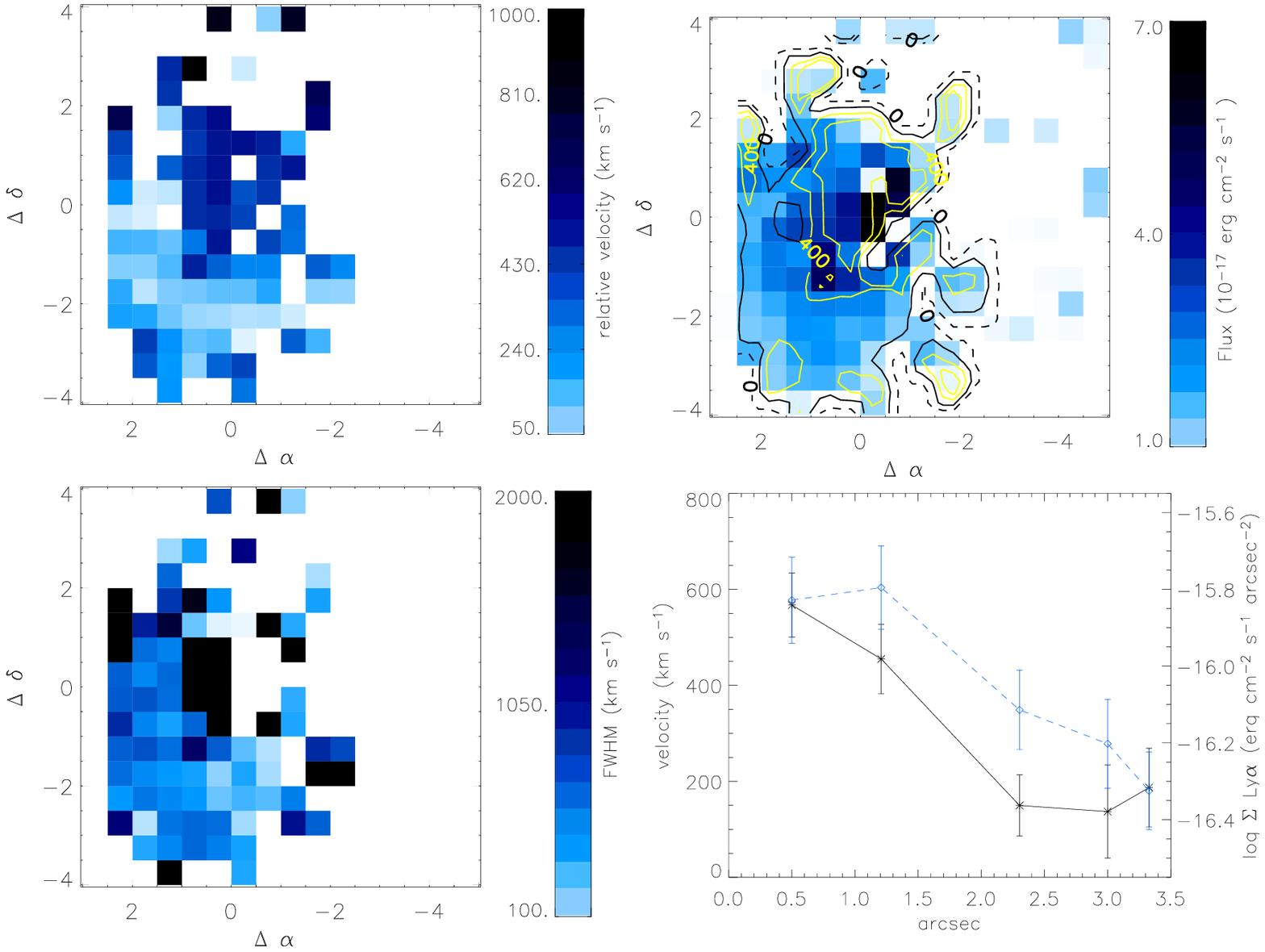}}
\caption{Maps of properties for the \lya\ emission nebula around
  Q1425+606.  \textit{Upper left panel}: Map of the \lya\ emission
  velocity offset relative to the QSO redshift.  \textit{Upper right
    panel}: Map of the integrated flux from each spaxel with smoothed
  velocity contours overlaid.  White fields correspond to spaxels
  where the \lya\ emission line could not be recognised visually,
  which therefore warps the velocity contours close to the QSO center
  at (0,0) where residuals are present.  \textit{Lower left panel}:
  Map of the \lya\ \textit{FWHM} corrected for the instrument
  resolution. The \textit{lower right panel} shows the radial profiles
  of the velocity component (solid line, left side axis) and
  \lya\ surface flux density (dashed line, right axis) starting from
  the center of the QSO along a direction with PA = 135$^{\circ}$ east
  of north. A fit of the radial surface brightness profile by an
  exponential function gives a scale length of $\sim$2\farcs5
  (20~kpc). \emph{[See the online edition of the Journal for a colour
      version of this figure.]}}
\label{fig:qlya_vel_q1425}
\end{figure*}

We start with the emission line nebula around Q1425+606.  For each
spaxel we fit a Gaussian profile to the spectrum around the expected
position of the \lya\ line using {\tt ngaussfit} in IRAF.  After an
automatic fit to all spectra in the cube, the fit of each individual
spectrum is checked interactively.  No constraints on the parameters
in the fits are imposed.  These fits are used to estimate for each
spaxel the line flux, centroid and \textit{FWHM}.  The \textit{FWHM}
is corrected for instrument resolution before deriving the velocities.
Results from these fits are shown in Fig.~\ref{fig:qlya_vel_q1425} and
are presented as maps that retain the individual spaxels rather than
the interpolated images shown in Fig.~\ref{fig:qlya_qso_lya}. Only
spaxels that have sufficient signal to allow for a visual detection of
an emission line are included. The upper left panel shows the relative
velocity offsets where as a reference point we use the QSO redshift
$z_{\mathrm{em}}=3.203$. We find evidence for velocities of $\sim$600
km s$^{-1}$ close to the QSO decreasing to 100--200 km~s$^{-1}$ at a
distance of 3\farcs5 from the QSO. In projection this corresponds to
$\sim25$~kpc. Using the redshifts inferred from the \ion{C}{iv} and
\ion{C}{ii} lines implies velocity offsets of $\sim$3000 km s$^{-1}$
at the center.

The surface brightness of the \lya\ emission ranges from
2$\times10^{-16}$ \ecs arcsec$^{-2}$ close to the QSO to about
2$\times10^{-17}$ \ecs arcsec$^{-2}$, 3\arcsec\ from the QSO.  After
correcting the measured \textit{FWHM} for the average instrumental
resolution, we find typical line widths of 10--15~{\AA} corresponding
to a velocity dispersion of 600--900~km~s$^{-1}$ as illustrated in the
lower left panel.  The \textit{FWHM} is slightly lower to the south of
the QSO (300~km~s$^{-1}$). There are indications of a high velocity
dispersion closer to the QSO, but the uncertainty is high in this
region due to QSO PSF subtraction residuals. In the upper right panel
in Fig.~\ref{fig:qlya_vel_q1425} we show the flux in each spaxel
overlayed by velocity contours. The lower right hand panel in
Fig.~\ref{fig:qlya_vel_q1425} shows the one-dimensional trace of the
velocity offset and \lya\ surface brightness distribution as points
connected by solid and dashed lines, respectively. These profiles are
derived at PA = 135$^{\circ}$ east of north where the emission line
region appears to have the largest extension.  Fluxes and velocities
at each point are derived from an average of four spaxels.  The
uncertainties are estimated from the standard deviations of these four
values which are combined in quadrature with typical uncertainties for
a single spaxel. The surface brightness profile has an exponential
scale length of $\sim$2\farcs5 which corresponds to 20 kpc at the QSO
redshift.

\begin{figure*}[!htpb]
\centering
\resizebox{\hsize}{!}{\includegraphics[bb= 40 400 555 795,clip]{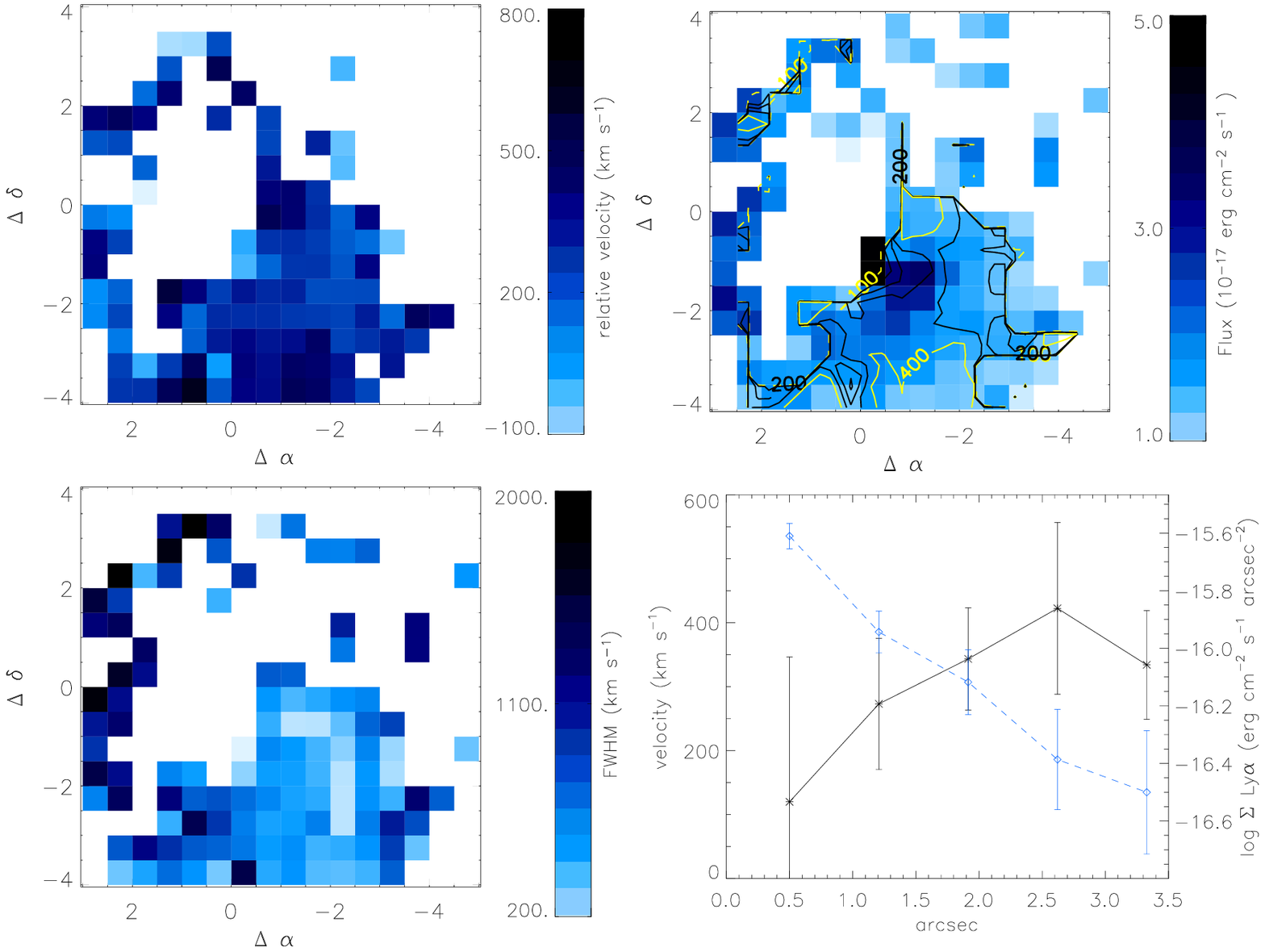}}
\caption{Maps of the \lya\ emission nebula surrounding Q1759+7539. The
  panels are similar to those described in
  Fig.~\ref{fig:qlya_vel_q1425}.  Profiles in the lower right panel
  are derived at PA = 225$^{\circ}$ east of north. A fit of an
  exponential function to the radial surface brightness profile gives
  a scale length of $\sim$1\farcs4 corresponding to 10~kpc at the QSO
  redshift. \emph{[See the online edition of the Journal for a colour
      version of this figure.]}  }
\label{fig:qlya_vel_q1759}
\end{figure*}

Maps of the velocity structures of the \lya\ nebula around Q1759+7539
are presented in Fig.~\ref{fig:qlya_vel_q1759}. The surface brightness
shows bright parts to the south-west of the QSO, but also to the east
some fainter emission appears. However, this is not clearly visible in
the on--off band image in Fig.~\ref{fig:qlya_qso_lya}. Including the
eastern part of the nebula, the total extension of the nebula appears
to be about 8\arcsec\ along the longest axis. Considering the bright
part of the nebula along PA~=~225$^{\circ}$, the velocity profile is
consistent with a slope of zero within 1 $\sigma$ uncertainties.  The
interpretation is difficult because of the complex velocity structure
in the nebula as shown by the contours in the upper right panel.  This
is also the case for the \textit{FWHM} in the lower left panel, but it
appears to be constant within errors at $\sim$450~km~s$^{-1}$ in the
bright part of the nebula.  The eastern part of the nebula shows
similar velocity offsets of 200$\pm$150~km~s$^{-1}$ relative to the
QSO redshift but the \textit{FWHM} is significantly larger:
$>1100$~km~s$^{-1}$.

The surface brightness profile has an exponential scale length of
1\farcs4 corresponding to $\sim$10~kpc. This is smaller by a factor of
two compared to the value estimated for the Q1425+606 nebula.

%________________________________________________________________

\section{Results}
\label{sect:qlya_scaling}
We detect extended \lya\ emission for 5 of the 7 systems. It is
possible that all 7 QSOs have extended emission, but one QSO is too
faint, and the other is affected by a CCD defect at the wavelength of
interest.  Table~\ref{tab:qlya_results} lists the flux, \textit{FWHM}
and extension for each detected EELR.  Fluxes are corrected for
galactic reddening using the dust-maps of \citet{schlegel98}.
Compared to the broad \lya\ emission lines from the QSO broad line
region, which have velocities around 10000 km~s$^{-1}$, the nebulae
have much narrower \lya\ emission lines ($\sim$500 km~s$^{-1}$).  The
emission appears to be asymmetric and mostly one-sided, and the
brightest extended \lya\ emission nebulae are found around the
brightest QSOs.

Only for the two brightest objects can the surface brightness be
measured.  The angular sizes of the brightest two nebulae are measured
along the long axis. For the fainter \lya\ emission regions the
extension is estimated by the offset between the QSO centroid and the
most distant emission seen in the IFS data. Errors of these values are
comparable to the spaxel size, which at the QSO redshifts sample
approximately 5 kpc.  The uncertainties for the velocity offsets
listed in column 9 are mainly caused by uncertainties in the
wavelength calibration leading to uncertainties in the derived
redshifts of $\Delta z\approx 0.0005$.  Velocity offsets are derived
from the emission lines in the one-dimensional spectra, and therefore
can differ from the velocities derived from the analyses of the
kinematics in Sect.~\ref{sect:vel}.

\begin{table*}
\centering
\begin{footnotesize}
  \begin{tabular}{llllllllll}
   \hline \hline
   \noalign{\smallskip}
(1) &(2) &(3) &(4) &(5) &(6) &(7) &(8) &(9) \\
Name  &  $z$ (\lya) & V        & $\Sigma$ (\lya)   & size & $f_{\textrm{tot}}$    &   $\log L_{\mathrm{tot}}$  & \textit{FWHM} & $\Delta$ V\\
     &             & (km s$^{-1}$) &(\ecs arcsec$^{-2}$)&(kpc)&  (10$^{-16}$~\ecs)  & (erg~s$^{-1}$) &  (km s$^{-1}$) &  (km s$^{-1}$)\\
  \noalign{\smallskip}
   \hline
  \noalign{\smallskip}
Q0953+4749  & 4.489 &          &                  & 13 &
   $0.36\pm0.17$ & 42.9 &  1000 & 1800$\pm$200 \\
   \noalign{\smallskip}
Q1425+606   & 3.204 & 600--200 &  2$\times10^{-16}$ & 34 &
   $9.8\pm0.8$ & 43.9 & 500  & 100$\pm$100\\ 
   \noalign{\smallskip}
Q1451+122   & 3.253 &          &                   & 15 &
   $1.8\pm0.5$ & 43.2 & 500 & --600$\pm$100\\
   \noalign{\smallskip}
Q1759+7539  & 3.049 & 200--300 &  3$\times10^{-16}$ & 60 & 
   $9.9\pm1.6$ & 43.9 & 450& 0$\pm$100\\
   \noalign{\smallskip}
Q2233+131   & 3.301 &          &                    & 10 &
   $1.1\pm0.4$ & 43.0 &  $<$400& 700$\pm$100\\
   \noalign{\smallskip}
\hline
   \noalign{\smallskip}
 \end{tabular}
  \caption[]{Properties of the \lya\ nebulae around the QSOs. Columns
  (1) list the names, (2) the redshifts derived from the narrow
  emission lines, (3) velocity offsets in the EELRs relative to the
  QSO redshifts, (4) peak surface brightness, (5) apparent extension
  of the \lya\ nebula, (6) total \lya\ flux from the nebula, (7) total
  luminosity in the nebulae, (8) Emission line \textit{FWHM} corrected
  for instrumental resolution, (9) Relative velocity offsets between
  the narrow emission lines (one-dimensional spectra) and the QSO
  redshifts. Fluxes have been corrected for Galactic reddening. An
  analysis of the surface brightness has been attempted only for the
  two objects with very bright nebulae.  }
  \label{tab:qlya_results}
\end{footnotesize}
\end{table*}

\subsection{Emission line fluxes}

How much flux is emitted by the EELRs compared to the overall flux of
the QSOs themselves?  The scalings of the spectra shown in
Fig.~\ref{fig:qlya_qso_lya} could indicate some relation.  To estimate
the flux in the IFS data, we integrate the extended narrow-\lya\ flux
in the nebulae from Gaussian fits to the one-dimensional spectra.  To
calculate the flux from the QSOs in the same wavelength interval we
integrate the one-dimensional QSO spectra from --10~{\AA} to +10~{\AA}
centered on the narrow \lya\ emission wavelength, and correct for
Galactic extinction as well.  We find that the flux densities in the
EELRs represent approximately 1--2\% of the QSO fluxes within the
narrow-band filter.

This fraction is much lower than that commonly found for RLQs.  In a
narrow-band imaging study of high redshift steep-spectrum,
lobe-dominated RLQs, \citet{heckman91b} (hereafter H91b) found that
extended \lya\ emission is a common feature around these objects. The
\lya\ EELRs contained a flux fraction of $\sim$10--25\% that of the
QSOs within the filter width of 15~{\AA} in the H91b sample.  The
narrow-band \lya\ flux from the QSOs spans the same range in the two
samples, but our data have \lya\ EELR fluxes approximately an order of
magnitude smaller than the H91b sample.  The band width of 20~{\AA}
that we choose corresponds roughly to the same rest-frame band width
for a $z=3.2$ system as a 15~{\AA} wide band at $z=2.2$, which is the
median redshift in the H91b sample. Hence the two samples are directly

H91b also estimates that the flux in the EELRs corresponds to about
10\% of the flux in the broad line region of the QSO taking into
account the flux falling outside the narrow-band filter. Likewise, if
we integrate the broad line \lya\ flux from the one-dimensional QSO
spectra, we find that the EELRs contain around 0.5\% of the flux
relative to the QSOs.

The median spatial extension in the H91b sample is 11\arcsec\ (or
90~kpc in the adopted cosmology), which is significantly larger than
the extensions found here. The two brightest objects have extensions
of $\sim$4\arcsec\ or $\sim$30 kpc at $z\approx3$.  The sensitivity in
the \citet{heckman91b} study is $1\times10^{-17}$ \ecs~arcsec$^{-2}$,
which is about a factor of two better than that reached by our IFS
observations.  Because of the cosmological dimming of $(1+z)^4$, the
redshift difference between the two samples has an effect on the
observed extension of the nebulae emission. The H91b sample has a
median redshift of 2.2, and their data therefore reach a factor of
\((1+3.11)^4/(1+2.2)^4=2.7\) fainter than ours (mean redshift 3.11).
Thus, the extension of the \lya\ nebulae around the RQQs could be 50\%
larger if the surface density profile is extrapolated to larger radii.
Even if we take into account this effect from the redshift
differences, and assume that the surface brightness profiles can be
extrapolated, the EELRs around RQQs have smaller sizes than those
around RLQs.

\subsection{Emission line luminosities}

Instead of the measured fluxes, we investigated whether the redshift
difference between the two samples has an effect and calculate the
luminosities for the given cosmology. For a given quasar luminosity,
the total luminosities in the \lya\ nebulae from RQQs are consistent
with being 10 times fainter relative to those from RLQs.

\begin{figure}
\centering
\resizebox{\hsize}{!}{\includegraphics{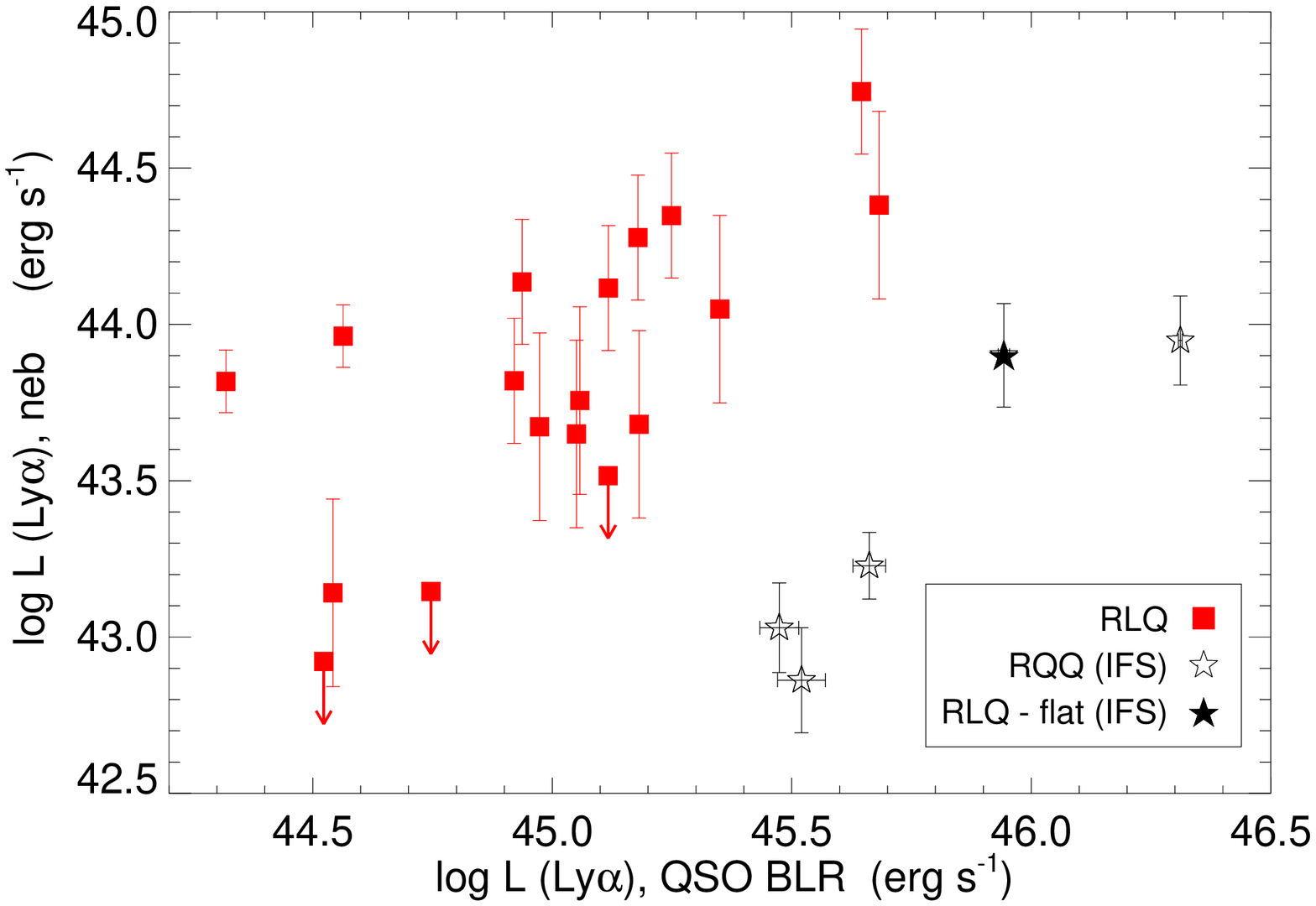}}
\caption{Luminosities in the EELRs as a function of the integrated
  luminosities for the QSO broad \lya\ lines.  The IFS results are
  indicated by stars, where the filled star represents the one
  core-dominated RLQ, and the outlined stars the four RQQs.  Squares
  represent the emission line nebulae surrounding 18 lobe-dominated
  RLQs from the sample in H91b. Arrows indicate upper limits for
  objects where no extended \lya\ emission was found.}
\label{fig:scaling_blr}
\end{figure}

In addition to the flux in the 20 {\AA} narrow-band image, we
investigate the fraction of the broad line region (BLR) line
luminosity for the QSOs relative to those for the EELRs. To calculate
the QSO \lya\ fluxes, we subtract the underlying powerlaw spectrum of
the QSO and integrate the one-dimensional spectra over the broad
\lya\ lines. The fluxes are corrected for Galactic extinction, and the
luminosities are calculated.  Fig.~\ref{fig:scaling_blr} shows the
luminosities of the nebulae as a function of the broad line region
luminosities. The broad \lya\ line fluxes for the RLQ are multiplied
by a factor of 2 to account for the flux that falls outside the
narrow-band images (see H91b). This figure shows that the luminosity
of the EELR contain about 0.5\% of the luminosities of the QSO broad
\lya\ lines.

Statistical tests support that there is a linear relation for the RQQs
and the one RLQ from the IFS sample in Fig.~\ref{fig:scaling_blr}. A
Pearson test gives a correlation coefficient of 0.92$\pm$0.04 for a
linear correlation, where the error bar is derived by boot strapping
techniques by adding random values corresponding to the uncertainties
for the luminosities.  Here the uncertainty of the QSO luminosities
are assumed to be dominated by photon shot noise.  A corresponding
analysis for the RLQs gives a moderate linear correlation coefficient
of 0.64$\pm$0.06, where the non detections in H91 are excluded.  For
the combined sample of 20 objects we find no obvious correlation, and
a correlation coefficient of --0.04$\pm$0.05.

To test whether there is a factor of 10 difference between the
luminosities of the \lya\ nebulae for the two samples, the EELR
luminosities from the IFS data are multiplied by 10. A Pearson test on
the total sample then gives a correlation coefficient of 0.71.  Hence,
the simple statistical tests justify the qualitative description of
the difference for the two samples given above.

It would be valuable to compare our study with the EELRs reported for
RQQs in the literature, but these are mainly based on slit spectra.
Therefore, a direct comparison cannot be done because the total flux
in the EELRs could be underestimated because of slit
losses. Nevertheless, the luminosity of one QSO and its \lya\ EELR in
\citet{steidel91} would place it very close to the two brightest
objects in our IFS study in Fig.~\ref{fig:scaling_blr}, and consistent
with the proposed scaling. The narrow-band image of Q1205--30 in
\citet{fynbo00b} and \citet{weidinger05} would place this object close
to the faintest objects, and still consistent with the correlation.
From a narrow-band image of one radio-weak QSO, \citet{bergeron99}
report a \lya\ EELR luminosity and extension similar to that of the
RLQs however, the narrow-band magnitude shows that the QSO is very
bright. We do not have sufficient information to evaluate whether the
properties of this QSO are consistent with any of the scaling
relations.

\subsection{Quasar luminosities}
\begin{figure}
\centering
  \resizebox{\hsize}{!}{\includegraphics{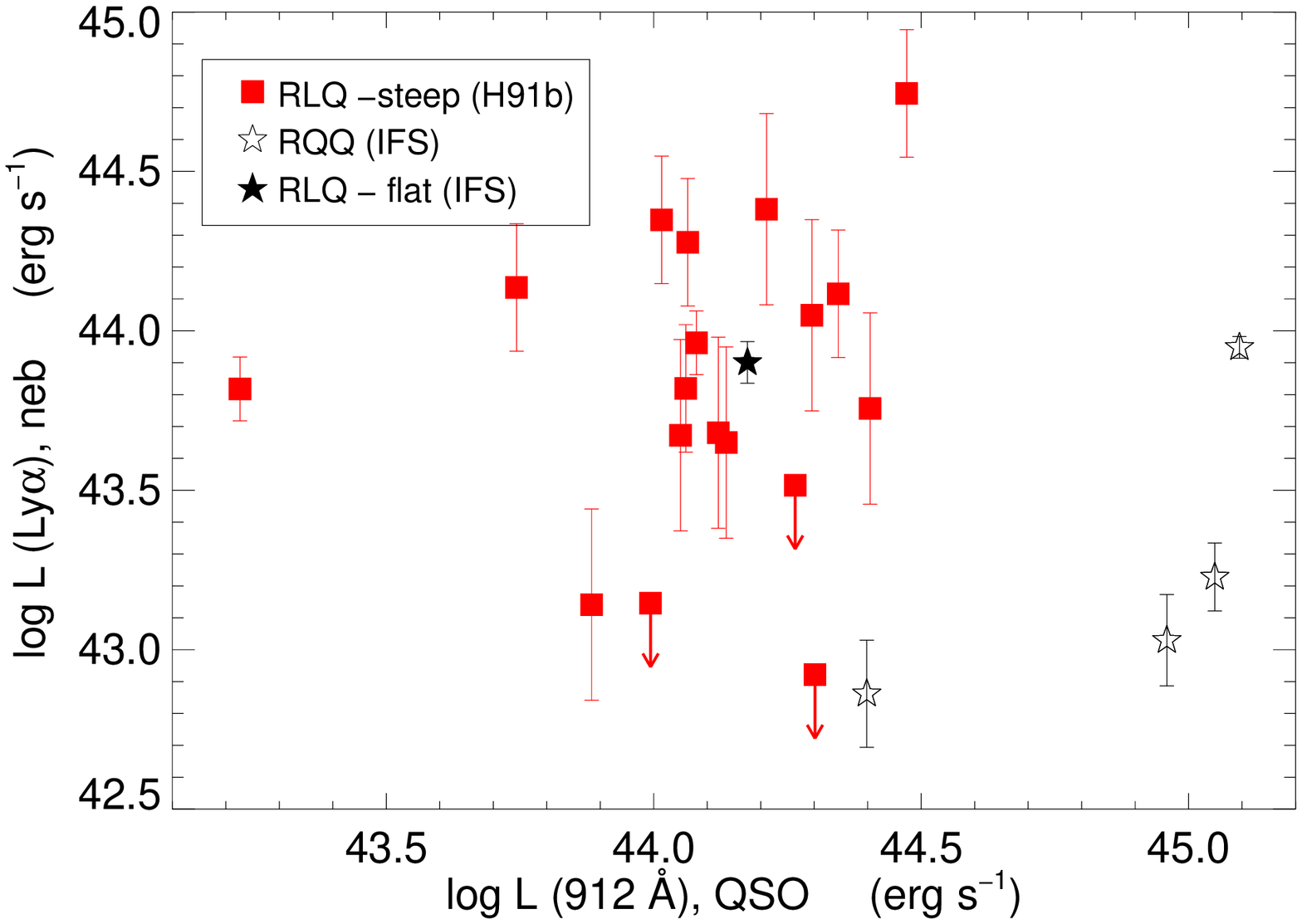}}
\caption{Extended \lya\ emission line luminosities as a function of
  the predicted quasar luminosities at the rest-frame 912~{\AA}. There
  is no significant correlation between the EELR luminosities and the
  nuclear ionising luminosities. }
\label{fig:qlya_abs1}
\end{figure}

To investigate whether the QSO ionising radiation has an effect on the
fluxes and extension of the EELRs, we first estimate the continuum
ionising flux from the QSOs. We take the $U,B$, or $V$ magnitudes
reported for the QSOs in the literature, and use a QSO template
spectrum to calculate a K-correction between the observed band and the
rest frame flux at 912~{\AA}.  The template spectrum is a hybrid of
the Sloan Digital Sky Survey composite spectrum \citep{vandenberk01}
and the composite FUSE spectrum \citep{scott04}.  For Q0953+4749, its
reported $V$ band magnitude is heavily affected by absorption in the
\lya\ forest. At $z=4.489$ the mean transmission in the \lya\ forest
is 0.33 \citep{songaila04}, thus we adopt a correction of 1.2
magnitudes in addition to the K-correction.  Most other quasars have
magnitudes measured slightly redwards of the \lya\ emission lines, or
the transmissions bluewards of the \lya\ lines are not as strongly
affected at $z\approx2$.  Fig.~\ref{fig:qlya_abs1} shows the
luminosities of the extended \lya\ emission line regions as a function
of the predicted quasar luminosities at 912~{\AA}.  Symbol shapes are
as in the previous figures. Taken at face value irrespective of the
small number statistics, the RQQs are offset to brighter values, but
this is a selection effect, because the quasars observed here are
bright and at higher redshifts. The one RLQ in the IFS sample has
properties similar to the RLQs in the H91b sample.

The figure suggests that the flux in the EELRs is independent of the
nuclear ionising emission, which is perhaps contrary to the
expectation. The same result is reached using K-corrections to
estimate the luminosities at rest-frame 1450 {\AA}, which shows that
the results are independent of the specific point of reference. Even
in the case that the ionising flux levels of the QSOs are estimated
incorrect by a factor of two due to the uncertainty in the individual
continuum levels relative to the template spectrum, this will not
affect the conclusion.

The flux we observe is the unabsorbed fraction. Possibly the results
are affected by dust extinction or absorption by neutral hydrogen in
the QSO surroundings. An analysis of $\sim$10000 QSO spectra have
revealed that reddening affects only a few QSOs \citep{hopkins04}, and
even very red QSOs show no evidence of dust obscuration
\citep{benn98}.

QSOs do generally not have Lyman limit edges
\citep{antonucci89,koratkar92}, which rules out significant amounts of
hydrogen absorption. Of the five QSOs studied here, two have no breaks
(Q0953+4749 and Q1425+606) and possibly also Q1759+7539, although our
spectrum only reaches 3650 {\AA}, i.e. 40 {\AA} bluewards of the
location of the break. The spectrum of Q2233+131 does not cover the
break. Q1451+122 shows indication of a break at 3840~{\AA}, i.e
bluewards by 3000~km~s$^{-1}$ from the QSO systemic redshift.  This
tells us that the absorption in the QSO lines of sight is not
significant, but the absorption towards EELRs cannot be quantified
with our data set. Assuming that the QSO sight lines are roughly
representative for the absorption, we find that neither extinction nor
absorption plays an important role.

\begin{figure}
\centering
\resizebox{\hsize}{!}{\includegraphics{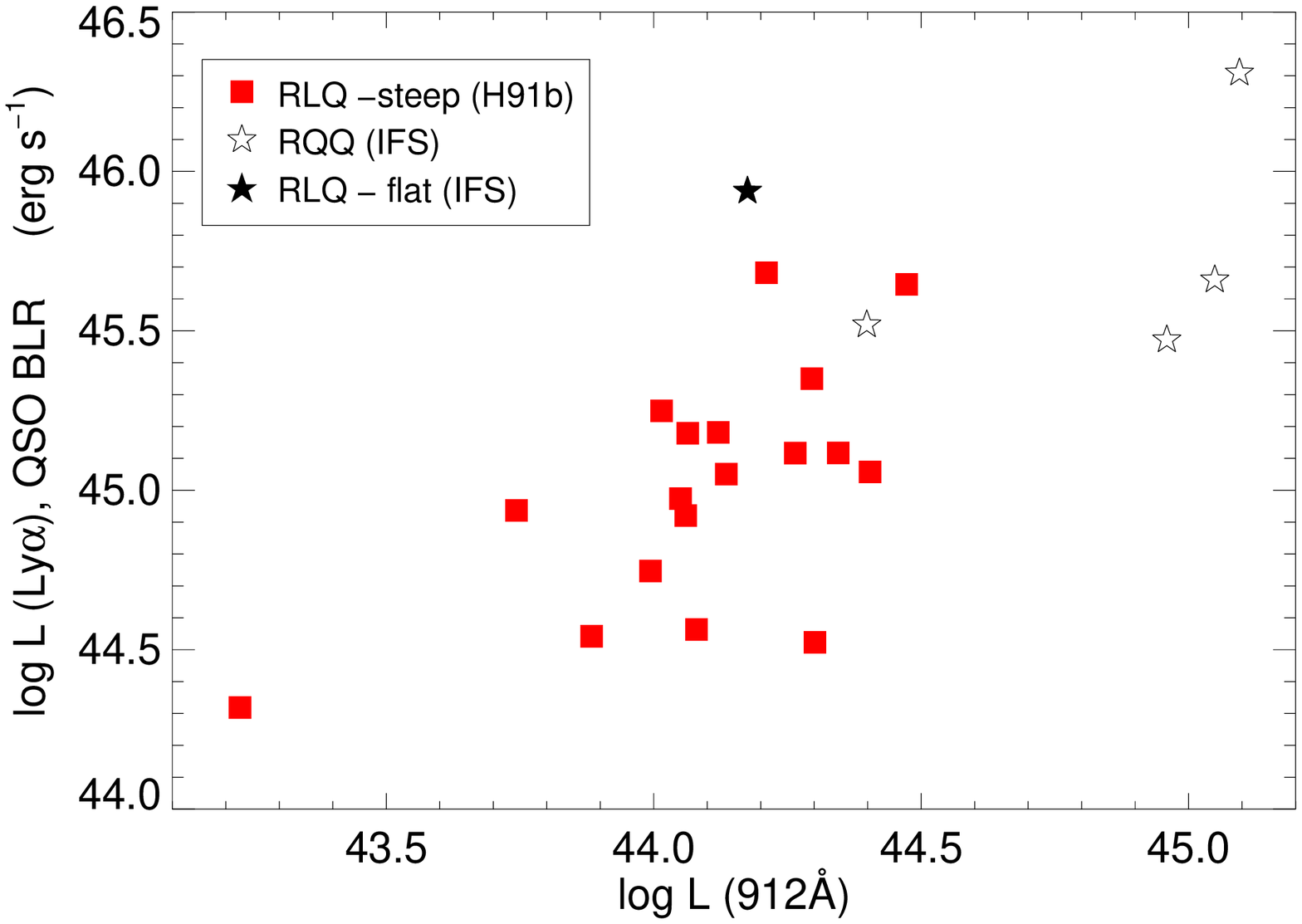}}
\caption{Integrated luminosities from the QSOs broad \lya\ lines as a
  function of the ionising flux calculated at 912 {\AA}.}
\label{fig:qso_lya912}
\end{figure}

Additionally we investigate how the quasar BLR \lya\ fluxes correlate
with the predicted quasar luminosities. As shown in
Fig.~\ref{fig:qso_lya912} there appears to be a correlation that
quasars with brighter 912~{\AA} continuum emission also have brighter
\lya\ emission. The total sample has a correlation coefficient
--0.72$\pm$0.05. Again Poissonian uncertainties are assumed and
bootstrap tests used to derive errors.  We will return to the
correlations in Sect.~\ref{sect:probs}.

\subsection{\lya\ nebula extension}
\begin{figure}[!t]
\centering  
  \resizebox{\hsize}{!}{\includegraphics{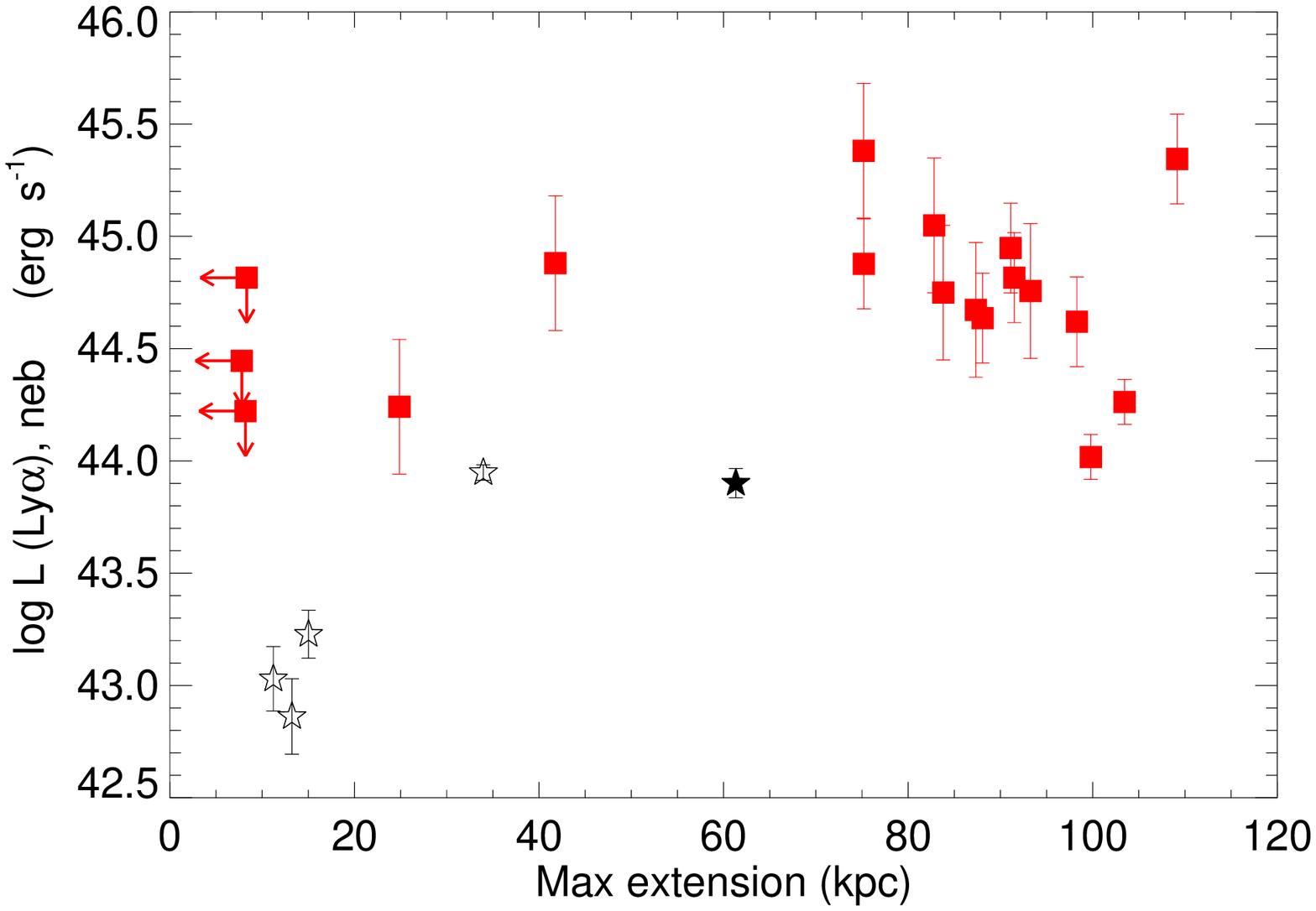}}
\caption{ Total \lya\ luminosity from the nebulae as a function of the
  maximum projected sizes.  Symbol shapes are as in the other figures
  and arrows indicate upper limits.}
\label{fig:qlya_scale4}
\end{figure}

The median size in the H91b sample is $\sim$90 kpc which is
significantly larger than the $\sim$30 kpc found for the brightest
nebulae in the IFS data. Fig.~\ref{fig:qlya_scale4} shows the total
\lya\ luminosity from the nebulae as a function of the apparent
maximum projected angular size. Smaller nebulae have lower
luminosities, whereas RLQs from H91b are larger and more luminous. A
Pearson test on the distribution of luminosities relative to the sizes
in the total sample gives a correlation coefficient of 0.44$\pm$0.03.

Morphologically there could be a difference between the RQQs and
RLQs. For the RQQs in this IFS study we find evidence for one-sided
emission only. Both one-sided emission \citep{weidinger05} and
two-sided \lya\ emission \citep{moller00} is reported for RQQs.  In
comparison, the \lya\ nebulae around the RLQs are asymmetric, and in
many cases two-sided emission is found (H91b).  Similarly, the one
core-dominated RLQ in Fig.~\ref{fig:qlya_vel_q1759} shows indications
of emission on both sides of the QSO.

 In the case of RLQ, the emission appearing asymmetric and two-sided
 could be explained by interactions of the jet with the ambient
 medium, since the brighter optical emission has been found aligned
 with the brighter radio emission (H91b). The asymmetry for the EELRs
 around RQQ must be explained by other means, for example by an
 intrinsic asymmetric distribution of material around the QSOs.

\subsection{EELR cloud masses}
Knowing the EELR luminosities and the Lyman continuum luminosity (at
1200~{\AA}), it is possible to estimate the masses of the hydrogen
clouds \citep[][equation 2c]{heckman91b} \(M\propto
L_{\mathrm{Ly\alpha}}U/Q\) M\subsun. We assume that the ionisation
parameter $U$ is the same for the RQQs studied here. The Lyman
continuum luminosity $Q$ is larger by a factor of a few for the RQQ in
our sample compared to the RLQ in H91b.  Since the EELR \lya\ luminosities
of the RQQs are smaller by a factor of $\sim$10, it follows that the
masses of the EELRs around RQQ are at least a factor of 10 smaller
than for RLQs, i.e. $10^7$--$10^8$~M\subsun.

The same conclusion would be reached if we follow the arguments in
\citet{villar-martin03}. Their equation to calculate gas cloud masses
around radio galaxies combines the luminosity $L$, volume of the cloud
$V$, and the filling factor $f$: \(M \propto (LVf)^{1/2}\). It is
difficult to estimate the volume of the clouds, but since the
extensions are smaller for the RQQ, the volumes are too.  Assuming a
similar filling factor would lead to masses in the range
$10^8$--$10^9$~M\subsun.

\subsection{General correlations for RQQ and RLQ}
\label{sect:probs}

The treatment of the correlations of the properties described above
avoided the upper limits measured for three of the RLQs.  Here we use
a generalised Kendall test \citep{isobe86}, which allows a treatment
of limits as well.  The total sample includes the 5 objects from our
IFS data, and the 18 objects in the H91b study.
Table~\ref{tab:qlya_kendall} reports the probabilities for the null
hypothesis that no correlation exists, i.e. small numbers are
equivalent to high probabilities. Only correlations giving
probabilities above 99\% are considered as indications of a
correlation.  The purpose is to find which fundamental relations
govern the observed properties of the emission line nebulae. For the
tests we use the total \lya\ luminosity in the EELRs
($L_{\mathrm{Ly\alpha,neb}}$), the total flux from the QSOs in the
broad \lya\ line ($L_{\mathrm{Ly\alpha,QSO}}$), the maximum extension
of the \lya\ nebulae (size), and the predicted quasar luminosities at
the rest frame 912~{\AA} ($L_{\mathrm{QSO},912})$.

The probabilities listed in Table~\ref{tab:qlya_kendall} are generally
quite high which indicates that no correlation is present. Two
exceptions exist. There is a probability of 99.7\% for a correlation
between the QSO \lya\ luminosity and the luminosities at 912~{\AA}.

Otherwise the luminosities at 912~{\AA} do not show strong
correlations with either nebulae sizes nor luminosities. This implies
that the ionising fluxes from the QSOs do not directly affect the
EELRs. If this is true, the main difference between the RLQ and RQQ
should lie in the environment.  The luminosity at 912~{\AA} is
uncertain due to the extrapolation of measured broad band magnitudes
with a QSO template spectrum. We investigate this effect by
multiplying the fluxes by a random factor between 0.5 and 2 and then
applying the correlation tests. We do not find changes in the
probabilities that significantly alter the conclusions.

The nebula sizes show the strongest correlation with the nebula
\lya\ luminosity with a probability of 99.9\% for a correlation. This
correlation implies that more luminous nebulae have larger sizes,
which is also shown in Fig.~\ref{fig:qlya_scale4}.

\begin{table}
\begin{footnotesize}
\centering
\begin{tabular}{l|lll}
\hline \hline 
\noalign{\smallskip}              
& $L_{\mathrm{Ly\alpha,neb}}$ & size & $L_{\mathrm{QSO},912\mathrm{\AA}}$ \\
\noalign{\smallskip}
\hline
\noalign{\smallskip}
$L_{\mathrm{Ly\alpha,QSO}}$& 0.174 (0.403)  &0.266 (--0.332) & 0.003 (0.885)\\
$L_{\mathrm{Ly\alpha,neb}}$& ...            &0.001 (1.020)   & 0.594 (--0.158)\\
size                  & ...            & ...            & 0.234 (--0.356)\\
  \noalign{\smallskip}
\hline
  \noalign{\smallskip}
\end{tabular}
\caption{Generalised Kendall's test probabilities for the null
  hypothesis that no relation exists between any two given quantities.
  The sample includes the IFS objects and the RLQs in H91b. The values
  in brackets give the correlation coefficients. The observed
  properties involved here are: the total luminosity in \lya\ in the
  EELRs ($L_{\mathrm{Ly\alpha,neb}}$), the flux from the QSOs in the
  broad \lya\ line ($L_{\mathrm{Ly\alpha,QSO}}$), the maximum
  extension of the \lya\ nebulae (size), the quasar luminosities
  estimated at the rest-frame 912~{\AA}
  ($L_{\mathrm{Ly\alpha,912{\AA}}}$).
}
\label{tab:qlya_kendall}
\end{footnotesize}
\end{table}

%--------------------------------------------------------------
\section{Summary and discussion}
\label{sect:qlya_disc}

We have presented evidence for differences in the extended
\lya\ nebulae around radio-quiet QSOs compared to radio-loud QSOs in
the literature.  We find that the brightest emission in these nebulae
is around 2--3$\times$10$^{-16}$~erg~cm$^{-2}$~s$^{-1}$~arcsec$^{-2}$
and extending to $\sim$4\arcsec.  Typical line widths are 500
km~s$^{-1}$, and the lines are shifted between --600 and 1800
km~s$^{-1}$ from the QSO systemic redshifts. The EELRs contain 1--2\%
of the QSO nuclear emission within the same pass band, or about 0.5\%
of the flux of the integrated broad \lya\ line from the QSOs.  These
are an order of magnitude smaller than found for RLQs studied in the
literature (H91b). Similarly, the nebula sizes analysed here are a
factor of a few smaller than detected around RLQs.  \cite{bremer92}
have suggested that the fainter EELRs around two RQQs were caused by a
smaller covering factor of the neutral material relative to
RLQs. Another interpretation is that the extended radio lobes, or
alternatively the same process that creates these, is responsible for
the main fraction of \lya\ line emission in steep-spectrum,
lobe-dominated RLQs.

Observations have shown that lobe-dominated RLQs have EELRs which are
aligned but not exactly correlated in location with the lobes
(H91b). Also optical emission line regions do not always show obvious
alignment with the radio emission \citep{crawford00}. This suggests
that the radio emission traces the direction of the AGN ionising cone,
but that an interaction between the radio-jet and the environment is
not necessarily the only cause. In some cases, jet interactions could
dominate the observed properties of the EELRs. For RQQs the properties
of the EELRs must be caused by a different mechanism.  The radio
emission is typically 2--3 orders of magnitude fainter than for
RLQs. If an interaction with a faint radio jet was solely responsible
for the \lya\ nebulae in the RQQs, the \lya\ emission would be much
fainter than that observed.

The AGN unification scheme states that the different radio properties
of radio-loud objects are related to their viewing angle
\citep{barthel89}, and core-dominated RLQs have jets very close to the
sight line.  Assuming that external conditions are similar for the two
bright emission nebulae studied here, a difference in viewing angle is
supported by the observed scale lengths in the surface brightness
profiles. The smaller scale-length of the nebula related to the
core-dominated RLQ would be consistent with the ionising emission from
the QSO having an orientation closer to the sight line than for the
RQQ.

Apart from jet interactions, other effects could determine the
properties and correlations we find for the EELRs.  Firstly, the
quasar ionising luminosity could be the main factor determining the
nebula luminosity. In this case it is expected that the continuum
luminosity from RLQs should be brighter than from RQQs, which is in
contradiction with the predicted 912~{\AA} luminosity. The statistical
tests give no evidence for a correlation of the line emission with the
quasar ionising luminosity.  Secondly, the velocities inferred from
the line widths of the \lya\ lines from the EELRs around RLQs could
suggest a disturbed medium relative to the EELRs around RQQs. In the
H91b sample of RLQs the nebular \lya\ line widths are
700--1000~km~s$^{-1}$, which is slightly larger than the $\sim$500 km
s$^{-1}$ found in this IFS study.  Radio galaxies show \lya\ line
fluxes similar to the bright objects in this sample and have velocity
dispersions of about 1000 km s$^{-1}$ \citep{vanojik97}, while the
quiescent component of the EELRs around radio-galaxies
\citep{villar-martin03} have velocities similar to those studied here.

Alternatively, external conditions can be responsible for the flux
differences.  One can suspect that the RQQs reside in less dense
environments \citep[see e.g.][]{vanojik97}. The electron density $n_e$
is proportional to the luminosity $L_{\mathrm{Ly\alpha}}$ divided by
the volume of the cloud $V$ times the covering factor $f$
\citep[i.e. \(n_e^2\propto \frac{L_{\mathrm{Ly\alpha}}}{fV}\),
  see][]{villar-martin03}. Since we find that both the volume and the
luminosity of the EELRs are smaller for RQQ, this would imply that the
density is roughly the same (around 10--100 cm$^{-3}$ for
$f=10^{-5}$). As we argue the masses for the RQQ nebulae are smaller,
this would imply that RQQ exist in less dense surroundings.

Nevertheless, observations of the host galaxies of the two populations
at low redshifts ($z<0.25$) indicate no differences for quasars with
similar nuclear powers \citep{dunlop03}. At higher redshifts little is
known about the environments of RQQs, but there is evidence that RLQs
reside in higher density environments
\citep[e.g.][]{hall98,sanchez99,sanchez02} compared to RQQs. Those
studies focused on the stellar light whereas we look at emission line
gas.  Rich environments are expected to have more gas by nature, and
by evolution. Captured galaxies suffer from a stripping effect due to
harassment and tidal effects, and the gas is ejected from the galaxies
into the intergalactic medium.  Such differences in the environment
could affect the \lya\ emission we detect. We know that the \lya\ flux
detected is a strict lower limit because of resonance scattering and
dust absorption.  We find evidence for resonance scattering in the
brightest emission lines that show red asymmetries.  This effect
may not affect the luminosity detected from the EELRs
significantly. The general lack of Lyman limit breaks in QSO spectra
places strong constraints on the amount of neutral gas present along
the sight lines.  This is in contrast to high redshift radio galaxies,
where 60\% of the galaxies show strong absorption lines that indicate
the presence of large amounts of neutral gas \citep{vanojik97}.

Extended \ion{C}{iv} and \ion{He}{ii} emission has been observed in
one RLQ which also has extended \lya\ emission \citep{lehnert98}, but
most RLQs only have upper limits for the \ion{C}{iv} line flux
\citep{heckman91}. For the two brightest objects considered here,
where the tightest constraints can be made, we find upper limits of
the flux ratio $f$(\ion{C}{iv})/$f$(\lya)$<$0.1. The limit is
comparable to the smallest values found for radio galaxies. A small
value can be caused by a difference in the viewing angle, or a low
ionisation parameter \citep{villar-martin96}.  Another explanation is
that the objects studied here have less metal enriched surroundings
\citep[see][]{weidinger05}, i.e. again suggesting that the environment
is different.

Finally, we must consider the possibility that the EELRs around the
RQQs are caused by infalling material which is ionised by the QSO. In
fact, the properties of the EELRs are similar to the flux and
extension calculated from theoretical predictions in the case of
matter falling into dark matter halos \citep{haiman01}. Assuming
infalling matter with a certain density profile, \citet{weidinger05}
modeled their observed surface brightness and velocity structure from
a long slit spectrum of Q1205--30. The structure of that nebula is
quite similar to the one-dimensional structure we extract for
Q1425+606. The interpretation of the nebulae structure and fluxes in
the infall scenario will depend strongly on the ambient density, and
variations in the nearby environment could cause differences such as
we find in the sample studied here.

\section{Conclusions}
\label{sect:conc}
Quasars are believed to reside in the most massive dark matter halos
at the centers of galaxies, and by studying quasars we can trace the
formation and evolution of massive galaxies at high redshifts.  The
growth of massive galaxies has been suggested to be regulated by
feedback from quasars. On the other hand, the formation of massive
galaxies require large amounts of gas to be present to form stars and
falling in to feed the quasar itself. Hence the evolution of massive
galaxies must be balanced between infall and feedback mechanisms, both
equally important for galaxy formation.

To investigate the impact of quasars on the surrounding gas, we have
observed extended emission from gas around very bright quasars at
$z\approx3$. We find indications for correlations between the
luminosities and sizes of EELRs around RQQs and compare them to those
already known for RLQs.  The scaling relations could indicate that the
interaction of the radio jets with the surrounding gas enhances the
luminosity of the EELRs around RLQs.  Since RQQ comprise about 90\% of
powerful AGN, the EELRs from RQQs seem more suitable to study the
effects of the central QSOs on the surrounding gas. Surprisingly, the
ionising fluxes at the Lyman limit for the QSOs themselves seem not to
determine the properties for the EELRs, since no correlations are
found, whereas there is a strong correlation between the EELR
luminosities and the integrated broad line \lya\ luminosity for the
five QSOs investigated. Why there is this discrepancy has yet to be
understood.

The sample studied here is very small, and correlations could be
biased because of small number statistics. Future investigations must
involve a well defined sample to examine RLQs and RQQs at similar
redshifts using the same technique to study the interaction between
the radio power, morphology and luminosities of the EELRs.

\begin{acknowledgements}
  L.~Christensen acknowledges support by the German Verbundforschung
  associated with the ULTROS project, grant no. 05AE2BAA/4. K. Jahnke
  acknowledges support from DLR project no.  50~OR~0404. We thank the
  referee M. Villar-Mart\'in for very thorough and detailed comments.
\end{acknowledgements}

\bibliography{ms5318}
\end{document}